\documentclass[10pt,tightenlines,eqsecnum,floats,aps,amsmath,amssymb,nofootinbib,prd,showpacs,superscriptaddress]{revtex4}
\usepackage{color}
\usepackage{amsmath}
\usepackage{amsfonts}
\usepackage{times}
\usepackage{graphics}
\usepackage{bm}
\usepackage[colorlinks, linkcolor=blue, anchorcolor=blue, citecolor=red]{hyperref}
\usepackage{subeqnarray}
\usepackage{cases}
\usepackage{bbm}
\usepackage{amssymb}
\usepackage{stmaryrd}
\usepackage{mathrsfs}

\usepackage{amsmath,amssymb,amsfonts}
\usepackage{graphicx}

\begin{document}

\title{Path Integrals and Alternative Effective Dynamics in Loop Quantum Cosmology}

\author{Li Qin}\email{qinli051@163.com}
\author{Guo Deng}\email{simba_dg@126.com}
\author{Yongge Ma\footnote{Corresponding author}}\email{mayg@bnu.edu.cn}

\affiliation{Department of Physics, Beijing Normal University, Beijing 100875, China}

\begin{abstract}
The alternative dynamics of loop quantum cosmology is examined by the path integral formulation. We consider the spatially flat FRW models with a massless scalar field, where the alternative quantization inherit more features from full loop quantum gravity. The path integrals can be formulated in both timeless and deparameterized frameworks. It turns out that the effective Hamiltonians derived from the two different viewpoints are equivalent to each other. Moreover, the first-order modified Friedmann equations are derived and predict quantum bounces for contracting universe, which coincide with those obtained in canonical theory.


\end{abstract}

\maketitle

\section{Introduction}
  In recent 25 years, considerable progress has been made in loop quantum gravity (LQG)\cite{Ro04,Th07,As04,Ma07}, which is a non-perturbative and background independent quantization of general relativity. A particular feature of LQG is to to inherit the core concept of  general relativity  that the fundamental physical theory should be background independent. The construction of LQG inspired the research on spin foam models (SFMs) \cite{Rovelli}, which are proposed as the path-integral formulation for LQG. In SFMs, the transition amplitude between physical quantum states is formulated as a sum over histories of all physically appropriate states. The heuristic picture is the following. One considers a 4-dimensional spacetime region M bounded by two spacelike 3-surfaces, which would correspond to the initial and final states for the path integral. The feature of SFMs is to employ the spin network states in LQG to represent the initial and final states. A certain path between the two boundaries is a quantum 4-geometry interpolated between the two spin networks. The interpolated spin foams can be constructed by considering the dual triangulation of M and coloring its surfaces with half integers and edges with suitable intertwiners. In order to obtain the physical inner product between the two boundary states, one has to sum over all possible triangulations and colorings \cite{Ro04}. This is certainly difficult to be achieved. Thus it is desirable to examine the idea and construction of SFMs by certain simplified models.

 It is well known that the idea and technique of canonical LQG are successfully carried out in the symmetry-reduced models, known as loop quantum cosmology (LQC) \cite{Bojowald}. Because of the homogeneity (and isotropy), the infinite degrees of freedom of gravity have been reduced to finite ones in LQC \cite{Boj00}. Hence LQC provides a simple arena to test the ideas and constructions of the full LQG. Therefore, it was proposed to test the idea and construction of SFMs by LQC models \cite{Henderson}. It was shown in Refs.\cite{spinfoam1,spinfoam2} that the transition amplitude in the deparameterized LQC of $k=0$ Friedmann universe equals the physical inner product in the timeless framework, and concrete evidence has been provided in support of the general paradigm underlying SFMs through the lens of LQC. How to achieve local spinfoam expansion in LQC was also studied in Refs. \cite{on the spinfoam, RV2}. Recently, the effective Hamiltonian constraint of $k=0$ LQC was derived from the path integral formulation in timeless framework in Ref.\cite{most recent}, and the effective measure introduces some conceptual subtleties in arriving at the WKB approximation. The path integrals and effective Hamiltonians of LQC for $k=+1,-1$ FRW models are also derived \cite{Huang}. The coherent state functional integrals are also proposed to study quantum cosmological models \cite{Qin}. However, in canonical LQC, there are some quantization ambiguities in constructing the Hamiltonian constraint operator. Theoretically, the quantization method which inherits more features from full LQG would be preferred. In spatially flat FRW model, two alternative versions of Hamiltonian constraint operator are proposed in Ref. \cite{YDM2}, where the big bang singularity resolution, correct classical limit and re-collapse of an expanding universe because of higher-order quantum effect are obtained.

 The purpose of this paper is to study the alternative effective dynamics of LQC from the viewpoint of path integral. The models which we are considering is the spatially flat FRW models with a massless scalar field. We will formulate the path integrals for two different proposed dynamics of LQC in both timeless and deparameterized frameworks. The multiple group averaging method proposed in Ref.\cite{Huang} is employed. It turns out that the alternative effective Hamiltonians derived from the two different viewpoints are equivalent to each other. Moreover, the first-order modified Friedmann equations can be derived, and the quantum bounces for contracting universe will also be obtained in both kinds of dynamics.
In section 2, we will introduce the basic frameworks of LQC and the path integral approach of multiple group averaging. Then we will derive the path integral formulation and effective Hamiltonian of alternative quantization model I and model II in sections 3 and 4 respectively. Finally a summary will be given in section 5.

\section{Basic Scheme}

We are consider the spatially flat FRW universe filled with a massless scalar field. In the kinematical setting, it is convenient to introduce an elementary cell ${\cal V}$ in the spatial manifold and restricts all integrations to this cell. Fixing a fiducial flat metric ${{}^o\!q}_{ab}$ and denoting by $V_o$ the volume of ${\cal V}$ in
this geometry, the physical volume reads $V=a^3V_o$ . The gravitational phase space variables ---the connections and the density weighted triads --- can be expressed as $ A_a^i = c\, V_o^{-(1/3)}\,\, {}^o\!\omega_a^i$ and $E^a_i = p\, V_o^{-(2/3)}\,\sqrt{{}^o\!q}\,\, {}^o\!e^a_i$, where $({}^o\!\omega_a^i, {}^o\!e^a_i)$ are a set of orthonormal co-triads and triads compatible with ${{}^o\!q}_{ab}$ and adapted to ${\cal V}$. $p$ is related to the scale factor $a$ via $|p|=V_o^{2/3}a^2$. The fundamental Poisson bracket is given by: $ \{c,\, p\} = {8\pi G\gamma}/{3} $, where $G$ is the Newton's constant and $\gamma$ the Barbero-Immirzi parameter. The gravitational part of the Hamiltonian constraint reads $C_{\mathrm{grav}} = -6 c^2\sqrt{|p|}/\gamma^2$.  To apply the area gap $\Delta$ of full LQG to LQC \cite{overview}, it is convenient to introduce variable $\bar{\mu}$ satisfying
\begin{align}
\label{f to mu}
\bar{\mu}^2~|p|=\Delta\equiv(4\sqrt{3}\pi\gamma)\ell_p^2,
\end{align}
where $\ell_p^2=G\hbar$, and new conjugate variables \cite{Robustness,CS,DMY,YDM2}:
\begin{align}
v:=\frac{\text{sgn}(p)|p|^{\frac{3}{2}}}{2\pi\gamma{\ell}^2_{\textrm{p}}\sqrt{\Delta}},
~~~~~b:=\frac{\bar{\mu}c}{2}.\label{v,b}
\end{align}
The new canonical pair satisfies $\{b,v\}=\frac{1}{\hbar}$. On the other hand, the matter phase space consists of canonical variables $\phi$ and $p_{\phi}$ which satisfy $\{\phi,p_{\phi}\}=1 $. To mimic LQG, the polymer-like representation is employed to quantize the gravity sector. The kinematical Hilbert space for gravity then reads $\mathcal{H}_{\rm{kin}}^{\rm{grav}}=L^2(\mathbb{R}_{\textrm{Bohr}},d\mu_H)$, where $\mathbb{R}_{\textrm{Bohr}}$ is the Bhor compactification of the real line and $d\mu_H$ is the Haar measure on it \cite{mathematical}. It turns out that the eigenstates of volume operator $\widehat{v}$, which are labeled by real number $v$, constitute an orthonomal basis in $\mathcal{H}_{\rm{kin}}^{\rm{grav}}$ as $\langle v_1| v_2\rangle=\delta_{v_1,v_2}$. For the scalar matter sector, one just uses the standard Schrodinger representation for its quantization, where the kinematical Hilbert space is $\mathcal{H}_{\rm{kin}}^{\rm{matt}}=L^2(\mathbb{R},d\phi)$. The total kinematical Hilbert Space of the system is a tensor product, $\mathcal{H}_{\rm{kin}}=\mathcal{H}_{\rm{kin}}^{\rm{grav}}\otimes\mathcal{H}_{\rm{kin}}^{\rm{matt}}$, of the above two Hilbert spaces. As a totally constrained system, the dynamics of this model is reflected in the Hamiltonian constraint $C_\mathrm{grav}+C_\mathrm{matt}=0$. Quantum mechanically, physical states are those satisfying quantum constraint equation
\begin{align}
(\widehat{C}_\mathrm{grav}+\widehat{C}_\mathrm{matt})\Psi(v,\phi)=0,\label{orin constraint}
\end{align}
which is not difficult to be rewritten as an Klein-Gordon like equation \cite{improved dynamics}:
\begin{align}
\widehat{C}\Psi(v,\phi)\equiv(\frac{\widehat{p}^2_{\phi}}{\hbar^2}-\widehat{\Theta})\Psi(v,\phi)=0. \label{constraint}
\end{align}
Eq.\eqref{constraint} indicates that we can get physical states by group averaging kinematical states as \cite{on the spinfoam}
\begin{align}
\Psi_f(v,\phi)=\lim\limits_{\alpha_o\rightarrow\infty}\int_{-\alpha_o}^{\alpha_o}
d\alpha ~e^{i\alpha\widehat{C}}~f(v,\phi),\quad\forall f(v,\phi)\in\mathcal{H}_{\rm{kin}},\label{group average}
\end{align}
and thus the physical inner product of two states reads
\begin{align}
\langle~f|g~\rangle_{\rm{phy}}=\langle
\Psi_f|g~\rangle=\lim\limits_{\alpha_o\rightarrow\infty}\int_{-\alpha_o}^{\alpha_o}
d\alpha ~\langle
f|e^{i\alpha\widehat{C}}|g\rangle.\label{innerproduct}
\end{align}
As is known, in timeless framework the transition amplitude equals to the physical inner product \cite{spinfoam1,spinfoam2}, i.e.,
\begin{align}
A_{tls}(v_f, \phi_f;~v_i,\phi_i)=\langle v_f,
\phi_f|v_i,\phi_i\rangle_{phy}=\lim\limits_{\alpha_o\rightarrow\infty}
\int_{-\alpha_o}^{\alpha_o}d\alpha\langle
v_f,\phi_f|e^{i\alpha\widehat{C}}|v_i,\phi_i\rangle.\label{amplitude}
\end{align}
On the other hand, Eq.\eqref{constraint} can also be written as
\begin{align}
\partial^2_\phi\Psi(v,\phi)+\widehat{\Theta}\Psi(v,\phi)=0.\label{Klein-Gorden}
\end{align}
The similarity between Eq.\eqref{Klein-Gorden} and Klein-Gorden equation suggests that one can regard $\phi$ as internal time, with respect to which gravitational field evolves.  In this deparameterized framework, we focus on positive frequency solutions, i.e., those satisfying
\begin{align}
-i\partial_\phi\Psi_+(v,\phi)=\widehat{\sqrt{\Theta}}\Psi_+(v,\phi)\equiv\widehat{\mathcal{H}}\Psi_+(v,\phi).
\label{positive frequency}
\end{align}
The transition amplitude in deparameterized framework is then given by
\begin{align}
A_{dep}(v_f,\phi_f;~v_i,\phi_i)=\langle v_f|e^{i\widehat{\mathcal{H}}(\phi_f-\phi_i)}|v_i\rangle,
\label{deparameterized amplitude1}
\end{align}
where $|v_i\rangle$ and $|v_f\rangle$ are eigenstates of volume operator in $\mathcal{H}_{\rm kin}^{\rm grav}$, and $\phi$ is the internal time.

 Starting with \eqref{amplitude}, we now consdier the transition amplitude of path integral under timeless framework. In order to compute $\langle v_f, \phi_f|e^{i\alpha\widehat{C}}|v_i,\phi_i\rangle$, a straightforward way is to split the exponential into $N$ identical pieces and insert complete basis as in \cite{most recent}. However, since $\alpha$ is the group-averaging parameter which goes from $-\infty$ to $\infty$, it is unclear whether $\alpha$ could be treated as the time variable $t$ in non-relativistic quantum mechanics path integral. We thus consider alternative path integral formulation of multiple group-averaging \cite{Huang}. Here one single group averaging (\ref{group average}) is generalized as multiple ones:
\begin{align}
&\lim\limits_{\alpha_o\rightarrow\infty}\int_{-\alpha_o}^{\alpha_o}d\alpha ~e^{i\alpha\widehat{C}}|v,\phi\rangle\nonumber\\
=&\lim\limits_{\tilde\alpha_{{No}},...,\tilde\alpha_{{1o}}\rightarrow\infty}\frac{1}{2\tilde\alpha_{{No}}}
\int_{-\tilde\alpha_\emph{{No}}}^{\tilde\alpha_\emph{{No}}}
d\tilde\alpha_N...\frac{1}{2\tilde\alpha_{{2o}}}\int_{-\tilde\alpha_{{2o}}}^{\tilde\alpha_{{2o}}}
d\tilde\alpha_2 \int_{-\tilde\alpha_{{1o}}}^{\tilde\alpha_{{1o}}}
d\tilde\alpha_1 ~e^{i(\tilde\alpha_1 +...+\tilde\alpha_N
)\widehat{C}}|v,\phi\rangle.
\end{align}
In order to trace the power for expansion, we re-scale the parameters by $\tilde\alpha_n=\epsilon\alpha_n$, where $\epsilon=\frac{1}{N}$. Then \eqref{amplitude} becomes
\begin{align}
&A_{tls}(v_f, \phi_f;~v_i,\phi_i)\nonumber\\
=&\lim\limits_{\alpha_\emph{{No}},...,\alpha_\emph{{1o}}\rightarrow\infty}\frac{1}{2\alpha_\emph{{No}}}
\int_{-\alpha_\emph{{No}}}^{\alpha_\emph{{No}}}
d\alpha_N...\frac{1}{2\alpha_\emph{{2o}}}\int_{-\alpha_\emph{{2o}}}^{\alpha_\emph{{2o}}}
d\alpha_2\cdot\epsilon
\int_{-\alpha_\emph{{1o}}}^{\alpha_\emph{{1o}}} d\alpha_1\langle
v_f,
\phi_f|e^{i\sum\limits_{n=1}^N{\epsilon\alpha_n}\widehat{C}}|v_i,\phi_i\rangle,
\label{amplitude2}
\end{align}
where $\alpha_\emph{no}=\tilde\alpha_\emph{no}/\epsilon, n=1,2,..,N$. Next, we are going to insert a set of complete basis at each knot.
Notice that $|v,\phi\rangle$ is the eigenstate of both volume operator and scalar operator simultaneously in $\mathcal{H}_{\rm kin}$, which is written as $|v\rangle|\phi\rangle$ for short,
and
\begin{align}
\mathbbm{1}_{\rm kin}=\mathbbm{1}_{\rm kin}^{\rm grav}\otimes\mathbbm{1}_{\rm kin}^{\rm matt}=\sum\limits_{v}|v\rangle\langle v|\int d\phi~|\phi\rangle\langle\phi|.
\end{align}
Thus, we have
\begin{align}
\langle v_f, \phi_f|e^{i\sum\limits_{n=1}^N{\epsilon\alpha_n}\widehat{C}}|v_i,\phi_i\rangle=\sum\limits_{v_{N-1},...v_1}\int d\phi_{N-1}...d\phi_1\prod\limits_{n=1}^N\langle \phi_n|\langle v_n|e^{i\epsilon\alpha_n\widehat{C}}|v_{n-1}\rangle\phi_{n-1}\rangle,
\label{insert basis}
\end{align}
where $v_f=v_N,\phi_f=\phi_N,v_i=v_0,\phi_i=\phi_0$ have been set. Since the constraint operator $\widehat{C}$ has been separated into gravitational part and material part, which live in $\mathcal{H}_{\rm kin}^{\rm grav}$ and $\mathcal{H}_{\rm kin}^{\rm matt}$ separately, we could calculate the exponential on each kinematical space separately. For the material part, one gets
\begin{align}
\langle{\phi_n}|e^{i\epsilon\alpha_n\frac{\widehat{p}^2_\phi}{\hbar^2}}|\phi_{n-1}\rangle
=&\int dp_{\phi_n}\langle{\phi_n}|p_{\phi_n}\rangle\langle p_{\phi_n}|e^{i\epsilon\alpha_n\frac{\widehat{p}^2_\phi}{\hbar^2}}|\phi_{n-1}\rangle\nonumber\\
=&\frac{1}{2\pi\hbar}\int dp_{\phi_n}e^{i\epsilon(\frac{p_{\phi_n}}{\hbar}\frac{\phi_n-\phi_{n-1}}{\epsilon}
+\alpha_n\frac{{p}^2_{\phi_n}}{\hbar^2})}.
\label{material amplitude}
\end{align}
As for the gravitational part, in the limit $N\rightarrow\infty(\epsilon\rightarrow0)$, the operator $e^{-i\epsilon\alpha_n \widehat{\Theta}}$ can be expanded to the first order, and hence we get
\begin{align}
\langle v_{n}|e^{-i\epsilon\alpha_n \widehat{\Theta}}|v_{n-1}\rangle=\delta_{v_n,v_{n-1}}-i\epsilon\alpha_n\langle v_{n}|\widehat{\Theta}|v_{n-1}\rangle+\mathcal{O}(\epsilon^2).
\label{piece}
\end{align}

\section{Path Integral and Effective Dynamics of Model I}

In this section, we employ one of the alternative Hamiltonian constraint operators for LQC proposed in \cite{YDM2}, where the \emph{extrinsic curvature} $K^{i}_{a}$ in the Lorentz term of the gravitational Hamiltonian constraint was quantized directly following the procedure in full LQG. Using the simplification treatment in \cite{Robustness}, we can get the action of the gravitational Hamiltonian operator in this model as:
\begin{align}
\widehat{\Theta}^{\rm F}|v\rangle&=\frac{3\pi G\gamma^2}{4}v
\big[(v+2)|v+4\rangle-2v|v\rangle+(v-2)|v-4\rangle\big]\nonumber\\
&\quad\quad-\frac{3\pi G(1+\gamma^2)}{16}v
\big[(v+4)|v+8\rangle-2v|v\rangle+(v-4)|v-8\rangle\big],\label{theta F}
\end{align}
which leads to
\begin{align}
\langle v_{n}|\widehat{\Theta}^{\rm F}|v_{n-1}\rangle&=\frac{3\pi G\gamma^2}{4}v_{n-1}\frac{v_{n}+v_{n-1}}{2}
(\delta_{v_{n},v_{n-1}+4}-2\delta_{v_{n},v_{n-1}}+\delta_{v_{n},v_{n-1}-4})\nonumber\\
 &\quad-\frac{3\pi G(1+\gamma^2)}{16}v_{n-1}\frac{v_{n}+v_{n-1}}{2}
(\delta_{v_{n},v_{n-1}+8}-2\delta_{v_{n},v_{n-1}}+\delta_{v_{n},v_{n-1}-8}).\label{matrix element1}
\end{align}
Applying \eqref{matrix element1} to \eqref{piece} and writing Kronecker delta as integral of $b_n$, which acts as the role of conjugate variable of $v_n$, we have
\begin{align}
&\langle v_{n}|e^{-i\epsilon\alpha_n \widehat{\Theta}_0}|v_{n-1}\rangle\nonumber\\
=&\frac{2}{\pi}\int^{\frac{\pi}{2}}_{0}db_n~e^{-ib_n(v_{n}-v_{n-1})}\Big(1-i\alpha_n\epsilon({3\pi G})v_{n-1}\frac{v_{n}+v_{n-1}}{2}\sin^2{(2b_n)}[1-(1+\gamma^2)\sin^2{(2b_n)}]\Big)+\mathcal{O}(\epsilon^2).\nonumber\\
\label{gravitational amplitude}
\end{align}
Applying \eqref{material amplitude} and \eqref{gravitational amplitude} to \eqref{amplitude2}, and then taking `continuum limit', we obtain
\begin{align}
&A^{\rm F}_{\rm tls}(v_f, \phi_f;~v_i,\phi_i)\nonumber\\
=&\lim\limits_{N\rightarrow\infty}~~~~\lim\limits_{\alpha_\emph{{No}},...,\alpha_\emph{{1o}}\rightarrow\infty}
\left(\epsilon\prod\limits_{n=2}^N\frac{1}{2\alpha_\emph{{no}}}\right)\int_{-\alpha_\emph{{No}}}^{\alpha_\emph{{No}}} d\alpha_N...\int_{-\alpha_\emph{{1o}}}^{\alpha_\emph{{1o}}} d\alpha_1\nonumber\\
&\times\int_{-\infty}^{\infty}d\phi_{N-1}...d\phi_1\left(\frac{1}{2\pi\hbar}\right)^N\int_{-\infty}^{\infty}
dp_{\phi_N}...dp_{\phi_1}\sum\limits_{v_{N-1},...,v_1}~\left(\frac{2}{\pi}\right)^N\int^{\frac{\pi}{2}}_{0}db_N...db_1\nonumber\\
&\times\prod\limits_{n=1}^{N}\exp{i\epsilon}\left[\frac{p_{\phi_n}}{\hbar}\frac{\phi_n-\phi_{n-1}}{\epsilon}
-{b_n}\frac{v_n-v_{n-1}}{\epsilon}+\alpha_n \left(\frac{p_{\phi_n}^2}{\hbar^2}-{3\pi G}v_{n-1}\frac{v_{n}+v_{n-1}}{2}
\sin^2{(2b_n)}[1-(1+\gamma^2)\sin^2{(2b_n)}]\right)\right].
\label{timeless F1}
\end{align}
Finally, we could write the above equation in path integral formulation as
\begin{align}
&A^{\rm F}_{\rm tls}(v_f, \phi_f;~v_i,\phi_i)\nonumber\\
=&c\int \mathcal{D}\alpha\int\mathcal{D}\phi\int\mathcal{D}p_{\phi}\int\mathcal{D}v\int\mathcal{D}b ~~\exp{\frac{i}{\hbar}\int_0^1d\tau \left[p_\phi\dot\phi-{\hbar}b\dot{v}+{\hbar}{\alpha}\left(\frac{p_\phi^2}{\hbar^2}-3\pi Gv^2\sin^2{(2b)}[1-(1+\gamma^2)\sin^2{(2b)}] \right)\right]}\label{timeless F2}
\end{align}
where $c$ is certain constant, and a dot over a letter denotes the derivative with respect to a \emph{fictitious time} variable $\tau$. The effective Hamiltonian constraint can be read out from Eq. (\ref{timeless F2}) as
\begin{align}
C^{\rm F}_{\rm eff}=\frac{p_\phi^2}{\hbar^2}-3\pi Gv^2\sin^2{(2b)}[1-(1+\gamma^2)\sin^2{(2b)}].  \label{effC}
\end{align}
Using this effective Hamiltonian constraint, we can explore the effective dynamics of universe by the modified Friedmann equation:
 \begin{equation}\label{Fdm eq1}
 H_{\rm F}^2=\frac{8\pi G\rho}{3}\left(1-\frac{\gamma^2+4(1+\gamma^2)\rho/\rho_{\rm c}}{1+\gamma^2}
 +\frac{\gamma^2\rho_{\rm c}}{2(1+\gamma^2)^2\rho}\left(1-\frac{4(1+\gamma^2)\rho}{\rho_{\rm c}}\right)
 \left(1-\sqrt{1-\frac{4(1+\gamma^2)\rho}{\rho_{\rm c}}}\right)\right),
 \end{equation}
where $\rho=\frac{p^2_{\phi}}{2V^2}$ is the matter density and $\rho_{\rm c}\equiv\frac{\sqrt3}{32\pi G^2\hbar\gamma^3}$ is a constant. This modified Friedmann equation coincides with the one in \cite{YDM2} if we ignore the higher-order quantum corrections therein. It is easy to see that if the matter density increase to $\rho=\frac{\rho_{\rm c}}{4(1+\gamma^2)}$, the Hubble parameter would be zero and the \emph{bounce} could occur for a contracting universe. On the other hand, in the classical region of large scale, we have $\rho\ll\rho_{\rm c}$ and hence Eq. (\ref{Fdm eq1}) reduces to the standard classical Friedmann equation: $H^2=\frac{8\pi G\rho}{3}$.

Besides the \emph{timeless} framework, we can also employ the above group-averaging viewpoint for the deparameterized framework. To this end, we define a new constraint operator $\widehat{C_+}=\frac{\widehat{p_{\phi}}}{\hbar}-\widehat{\mathcal{H}}$. Then Eq.\eqref{positive frequency} can be rewritten as
\begin{align}
\widehat{C_+}\Psi_+(v,\phi)=0.
\end{align}
The transition amplitude for this new Hamiltonian constraint reads
\begin{align}
A_{\rm dep}(v_f,\phi_f;~v_i,\phi_i)=\lim\limits_{\alpha_o\rightarrow\infty}\int_{-\alpha_o}^{\alpha_o}
d\alpha \langle v_f,\phi_f|e^{i\alpha
\widehat{C_+}}|v_i,\phi_i\rangle=\lim\limits_{\alpha_o\rightarrow\infty}\int_{-\alpha_o}^{\alpha_o}
d\alpha \langle
v_f,\phi_f|2\widehat{|{p}_\phi|}\widehat{\theta(p_{\phi})}e^{i\alpha
\widehat{C}}|v_i,\phi_i\rangle, \label{deparameterized amplitude2}
\end{align}
where
\begin{align}
\widehat{|{p}_\phi|}|p_{\phi}\rangle=|p_{\phi}||p_{\phi}\rangle,~~
\widehat{\theta(p_{\phi})}|p_{\phi}\rangle=
\begin{cases}
0& \text{$p_{\phi}\leq0$}\\
|p_{\phi}\rangle& \text{$p_{\phi}>0$}.
\end{cases}
\end{align}
Similar to the timeless case, the integration over single $\alpha$ can be written as multiple integrations,
\begin{align}
&A_{\rm dep}(v_f, \phi_f;~v_i,\phi_i)\nonumber\\
=&\lim\limits_{\alpha_\emph{{No}},...,\alpha_\emph{{1o}}\rightarrow\infty}\frac{1}{2\alpha_\emph{{No}}}
\int_{-\alpha_\emph{{No}}}^{\alpha_\emph{{No}}}
d\alpha_N...\frac{1}{2\alpha_\emph{{2o}}}\int_{-\alpha_\emph{{2o}}}^{\alpha_\emph{{2o}}}
d\alpha_2\cdot\epsilon\int_{-\alpha_\emph{{1o}}}^{\alpha_\emph{{1o}}}
d\alpha_1 ~~\langle v_f,
\phi_f|e^{i\sum\limits_{n=1}^N{\epsilon\alpha_n}\widehat{C_+}}|v_i,\phi_i\rangle.
\label{deparameterized amplitude3}
\end{align}
Since we work in the deparameterized case, it is reasonable to insert in the completeness relation \cite{two points}
\begin{align}
\mathbbm{1}=\sum\limits_v|v,\phi,+\rangle\langle v,\phi,+|,
\end{align}
where
\begin{align}
|v,\phi,+\rangle=\lim\limits_{\beta_o\rightarrow\infty}\int_{-\beta_o}^{\beta_o}
d\beta~e^{i\beta\widehat{C_+}}|v,\phi\rangle.\label{basis}
\end{align}
Note that the Hilbert space in deparameterized framework is unitarily equivalent to the physical Hilbert space with the basis \eqref{basis}.
Then the first piece of the exponential in \eqref{deparameterized amplitude3} becomes
\begin{align}
\langle v_1,\phi_1,+|v_{0},\phi_{0},+\rangle=&\lim\limits_{\alpha_{1o}\rightarrow\infty}\int_{-\alpha_{1o}}^{\alpha_{1o}}
d(\epsilon\alpha_1)~\langle
v_1,\phi_1,+|e^{i\epsilon\alpha_1\widehat{C_+}}|v_{0},\phi_{0}\rangle\nonumber\\
=&\lim\limits_{\beta'_{1o}\rightarrow\infty}\int_{-\beta'_{1o}}^{\beta'_{1o}}
d\beta'_1\langle
v_1,\phi_1|e^{i\beta'_1\widehat{C_+}}|v_{0},\phi_{0}\rangle\nonumber\\
=&\lim\limits_{\beta_{1o}\rightarrow\infty}\epsilon\int_{-\beta_{1o}}^{\beta_{1o}}
d\beta_1\langle
v_1,\phi_1|e^{i\epsilon\beta_1\widehat{C_+}}|v_{0},\phi_{0}\rangle,
\end{align}
and the last piece of the exponential reads
\begin{align}
&\lim\limits_{\alpha_{No},\beta'_{No}\rightarrow\infty}\frac{1}{2\alpha_{No}}\int_{-\alpha_{No}}^{\alpha_{No}}
 d\alpha_N\int_{-\beta'_{No}}^{\beta'_{No}} d\beta'_N~\langle v_N,\phi_N|e^{i\epsilon\alpha_N\widehat{C_+}}e^{i\beta'_N\widehat{C_+}}|v_{N-1},\phi_{N-1}\rangle\nonumber\\
=&\lim\limits_{\beta'_{No}\rightarrow\infty}\int_{-\beta'_{No}}^{\beta'_{No}}
d\beta'_N~\langle
v_N,\phi_N|e^{i\beta'_N\widehat{C_+}}|v_{N-1},\phi_{N-1}\rangle\nonumber\\
=&\lim\limits_{\beta_{No}\rightarrow\infty}\epsilon\int_{-\beta_{No}}^{\beta_{No}}
d\beta_N~\langle
v_N,\phi_N|e^{i\epsilon\beta_N\widehat{C_+}}|v_{N-1},\phi_{N-1}\rangle.
\end{align}
The remaining pieces of the exponential can also be expressed as
\begin{align}
&\lim\limits_{\alpha_{no}\rightarrow\infty}\frac{1}{2\alpha_{no}}\int_{-\alpha_{no}}^{\alpha_{no}}
d\alpha_n~\langle v_n,
\phi_n,+|e^{i\epsilon\alpha_n\widehat{C_+}}|v_{n-1},\phi_{n-1},+
\rangle\nonumber\\
=&\lim\limits_{\beta'_{no}\rightarrow\infty}\int_{-\beta'_{no}}^{\beta'_{no}}
d\beta'_n~\langle
v_n,\phi_n|e^{i\beta'_n\widehat{C_+}}|v_{n-1},\phi_{n-1}\rangle\nonumber\\
=&\lim\limits_{\beta_{no}\rightarrow\infty}\epsilon\int_{-\beta_{no}}^{\beta_{no}}
d\beta_n~\langle
v_n,\phi_n|e^{i\epsilon\beta_n\widehat{C_+}}|v_{n-1},\phi_{n-1}\rangle.
\end{align}
So \eqref{deparameterized amplitude3} becomes
\begin{align}
&A_{\rm dep}(v_f, \phi_f;~v_i,\phi_i)\nonumber\\
=&\lim\limits_{\beta_{{No}},...,\beta_{{1o}}\rightarrow\infty}\epsilon^{N}
\int_{-\beta_{{No}}}^{\beta_{{No}}}
d\beta_N...\int_{-\beta_{{1o}}}^{\beta_{{1o}}}
d\beta_1\sum\limits_{v_{N-1},...,v_1}
\prod\limits_{n=1}^N~\langle v_n, \phi_n|e^{i\epsilon\beta_n\widehat{C_+}}|v_{n-1},\phi_{n-1}\rangle.\nonumber\\
=&\lim\limits_{\beta_\emph{{No}},...,\beta_\emph{{1o}}\rightarrow\infty}\epsilon^{N}
\int_{-\beta_\emph{{No}}}^{\beta_\emph{{No}}}
d\beta_N...\int_{-\beta_\emph{{1o}}}^{\beta_\emph{{1o}}}
d\beta_1\sum\limits_{v_{N-1},...,v_1}\prod\limits_{n=1}^N~\langle
v_n,
\phi_n|2\widehat{|{p}_{\phi_n}|}\widehat{\theta({p}_{\phi_n})}e^{i\epsilon\beta_n\widehat{C}}|v_{n-1},\phi_{n-1}
\rangle,\label{damplitude}
\end{align}
where \eqref{deparameterized amplitude2} is applied in second step.
Now we can split each piece in \eqref{damplitude} into gravitational and material parts.
Calculations similar to those in timeless framework lead to
\begin{align}
&A^{\rm F}_{\rm dep}(v_f, \phi_f;~v_i,\phi_i)\nonumber\\
=&\lim\limits_{N\rightarrow\infty}~~~~\lim\limits_{\beta_\emph{{No}},...,\beta_\emph{{1o}}\rightarrow\infty}\epsilon^{N}
\int_{-\beta_\emph{{No}}}^{\beta_\emph{{No}}}
d\beta_N...\int_{-\beta_\emph{{1o}}}^{\beta_\emph{{1o}}}
d\beta_1\left(\frac{1}{2\pi\hbar}\right)^N\int_{-\infty}^{\infty}dp_{\phi_N}...dp_{\phi_1}\sum
\limits_{v_{N-1},...,v_1}~~~\left(\frac{2}{\pi}\right)^N\int^{\frac{\pi}{2}}_{0}db_N...db_1\nonumber\\
&\times\prod\limits_{n=1}^{N}2|p_{\phi_n}|\theta(p_{\phi_n})\exp{i\epsilon}\Big[\frac{p_{\phi_n}}{\hbar}
\frac{\phi_n-\phi_{n-1}}{\epsilon}-{b_n}\frac{v_n-v_{n-1}}{\epsilon}\nonumber\\
&\quad\quad\quad\quad\quad\quad\quad\quad\quad\quad+\beta_n\Big(\frac{p_{\phi_n}^2}{\hbar^2}-{3\pi G}v_{n-1}\frac{v_{n}+v_{n-1}}{2}
\sin^2{(2b_n)}[1-(1+\gamma^2)\sin^2{(2b_n)}]\Big)\Big].
\end{align}
We can integrate out $\beta_n$ and $p_{\phi_n}$ and arrive at
\begin{align}
&A^{\rm F}_{\rm dep}(v_f, \phi_f;~v_i,\phi_i)\nonumber\\
=&\lim\limits_{N\rightarrow\infty}
\sum\limits_{v_{N-1},...,v_1}~~~\left(\frac{2}{\pi}\right)^N
\int^{\frac{\pi}{2}}_{0}db_N...db_1\nonumber\\
&\times\prod\limits_{n=1}^{N}\exp{i\epsilon(\phi_f-\phi_i)}\left[\sqrt{{3\pi G}v_{n-1}\frac{v_{n}+v_{n-1}}{2}\sin^2{(2b_n)}[1-(1+\gamma^2)\sin^2{(2b_n)}]}-{b_n}
\frac{v_n-v_{n-1}}{\epsilon(\phi_f-\phi_i)}\right]\nonumber\\
=&c'\int\mathcal{D}v\int\mathcal{D}b~\exp\frac{i}{\hbar}\int d\phi(\sqrt{3\pi G\hbar^2v^2\sin^2{(2b)}[1-(1+\gamma^2)\sin^2{(2b)}]}
-{\hbar\dot{v}b}),\label{damplitude2}
\end{align}
where $\dot{v}=\frac{dv}{d\phi}$. The effective Hamiltonian in deparameterized framework can be read out from \eqref{damplitude2} as: \begin{equation}
\mathcal{H}^{\rm F}_{\rm eff}=-\sqrt{3\pi G\hbar^2v^2\sin^2{(2b)}[1-(1+\gamma^2)\sin^2{(2b)}]},\label{effc H}
\end{equation}
We can use this effective Hamiltonian to get the effective equations of motion:
\begin{eqnarray}
&&\dot{v}=\frac{3\pi Gv^2\sin{(4b)}[1-2(1+\gamma^2)\sin^2{(2b)}]}
{\sqrt{3\pi Gv^2\sin^2{(2b)}[1-(1+\gamma^2)\sin^2{(2b)}]}},\\
&&\dot{b}=-\textrm{sgn}(v)\sqrt{3\pi {G} v^2\sin^2{(2b)}[1-(1+\gamma^2)\sin^2{(2b)}]},
\end{eqnarray}
which also predict a \emph{bounce} of a contracting universe when $\dot{v}=0$. It is easy to see that the bounce point coincides with that from \eqref{Fdm eq1}. In fact, the effective equations derived in timeless and deparameterized frameworks coincides with each other. Hence, at least at first-order level, both methods of path integral confirm the effective dynamics of this model in canonical theory.

\section{Path Integral and Effective Dynamics of Model II}

In the other Hamiltonian constraint operator proposed in \cite{YDM2}, the Lorentz term was constructed by using the fact that the extrinsic curvature $K^i_a$ is related to the connection $A^i_a$ by $A^i_a=\gamma{K^i_a}$ in the spatially flat case. Reexpressing the connection by holonomy, we can get a simplified version of this operator $\hat{\Theta}^{\rm S}$ similar to $\hat{\Theta}^{\rm F}$ as:
\begin{align}
\widehat{\Theta}^{\rm S}|v\rangle&=\frac{3\pi G\gamma^2}{4}v
\big[(v+2)|v+4\rangle-2v|v\rangle+(v-2)|v-4\rangle\big]\nonumber\\
&\quad\quad-{3\pi G(1+\gamma^2)}v
\big[(v+1)|v+2\rangle-2v|v\rangle+(v-1)|v-2\rangle\big].\label{theta S}
\end{align}
Following the same procedure in the previous section, we can get the transition amplitude in timeless framework as:
\begin{align}
&A^{\rm S}_{\rm tls}(v_f, \phi_f;~v_i,\phi_i)\nonumber\\
=&\lim\limits_{N\rightarrow\infty}~~~~\lim\limits_{\alpha_\emph{{No}},...,\alpha_\emph{{1o}}\rightarrow\infty}
\left(\epsilon\prod\limits_{n=2}^N\frac{1}{2\alpha_\emph{{no}}}\right)\int_{-\alpha_\emph{{No}}}^{\alpha_\emph{{No}}} d\alpha_N...\int_{-\alpha_\emph{{1o}}}^{\alpha_\emph{{1o}}} d\alpha_1\nonumber\\
&\times\int_{-\infty}^{\infty}d\phi_{N-1}...d\phi_1\left(\frac{1}{2\pi\hbar}\right)^N\int_{-\infty}^{\infty}
dp_{\phi_N}...dp_{\phi_1}\sum\limits_{v_{N-1},...,v_1}~\frac{1}{\pi^N}\int^{{\pi}}_{0}db_N...db_1\nonumber\\
&\times\prod\limits_{n=1}^{N}\exp{i\epsilon}\left[\frac{p_{\phi_n}}{\hbar}\frac{\phi_n-\phi_{n-1}}{\epsilon}
-{b_n}\frac{v_n-v_{n-1}}{\epsilon}+\alpha_n \left(\frac{p_{\phi_n}^2}{\hbar^2}-{12\pi G}v_{n-1}\frac{v_{n}+v_{n-1}}{2}
\sin^2{b_n}(1+\gamma^2\sin^2{b_n})\right)\right],
\label{timeless S}
\end{align}
which gives an effective Hamiltonian constraint:
\begin{equation}
 C^{\rm S}_{\rm eff}=\frac{p^2_{\phi}}{\hbar^2}-12\pi{G}v^2\sin^2{b}(1+\gamma^2\sin^2{b}).
\end{equation}
From this effective constraint, we can derive another modified Friedmann equation:
 \begin{equation}
 H^2_{\rm S}=\frac{8\pi{G}\rho}{3}\left(1-\frac{3(1+\gamma^2)+\gamma^2\rho/\rho_{\rm c}}{\gamma^2}
 +\frac{2(1+\gamma^2)\rho_{\rm c}}{\gamma^4\rho}\left(\sqrt{1+\frac{\gamma^2\rho}{\rho_{\rm c}}}\left(1+\frac{\gamma^2\rho}{\rho_{\rm c}}\right)-1\right)\right).\label{Fdm eq2}
 \end{equation}
It is obvious that this effective equation is different from Eq.(\ref{Fdm eq1}) of the model I. By Eq.(\ref{Fdm eq2}), the quantum bounce would occur when matter density increase to $\rho=4(1+\gamma^2)\rho_{\rm c}$ for a contracting universe. It is easy to see that the Friedmann equation (\ref{Fdm eq2}) can also reduce to the classical one when $\rho\ll\rho_{\rm c}$. Moreover, Eq.(\ref{Fdm eq2}) coincides with the corresponding effective Friedmann equation in \cite{YDM2} if the higher-order quantum corrections therein are neglected.

Similarly, we can also get the deparameterized amplitude:
\begin{align}
&A^{\rm S}_{\rm dep}(v_f, \phi_f;~v_i,\phi_i)\nonumber\\
=&\lim\limits_{N\rightarrow\infty}
\sum\limits_{v_{N-1},...,v_1}~~~\frac{1}{\pi^N}
\int^{{\pi}}_{0}db_N...db_1\nonumber\\
&\times\prod\limits_{n=1}^{N}\exp{i\epsilon(\phi_f-\phi_i)}\left[\sqrt{{12\pi G}v_{n-1}\frac{v_{n}+v_{n-1}}{2}\sin^2{b_n}(1+\gamma^2\sin^2{b_n})}-{b_n}
\frac{v_n-v_{n-1}}{\epsilon(\phi_f-\phi_i)}\right]\nonumber\\
=&c'\int\mathcal{D}v\int\mathcal{D}b~\exp\frac{i}{\hbar}\int d\phi(\sqrt{12\pi G\hbar^2v^2\sin^2{b}(1+\gamma^2\sin^2{b})}
-{\hbar\dot{v}b}),\label{damplitude3}
\end{align}
which gives an effective Hamiltonian:
\begin{equation}
\mathcal{H}^{\rm S}_{\rm eff}=-\sqrt{12\pi G\hbar^2v^2\sin^2{b}(1+\gamma^2\sin^2{b})}.\label{effc H2}
\end{equation}
The evolution of $v$ and $b$ with respect to $\phi$ can be obtained from this effective Hamiltonian as:
\begin{eqnarray}
&&\dot{v}\equiv\frac{dv}{d\phi}=\frac{12\pi{G}v^2\sin{(2b)}(1+2\gamma^2\sin^2{b})}{2\sqrt{12\pi G\hbar^2v^2\sin^2{b}(1+\gamma^2\sin^2{b})}},\nonumber\\
&&\dot{b}\equiv\frac{db}{d\phi}=-\textrm{sgn}(v)\sqrt{12\pi G\hbar^2v^2\sin^2{b}(1+\gamma^2\sin^2{b})},
\end{eqnarray}
which again coincide with the effective Eq.(\ref{Fdm eq2}) and predict the same bounce when $\dot{v}=0$. Hence, the corresponding effective dynamics of this model in canonical theory is also confirmed at first order by the path integral.

\section{Summary}

The motivations to study alternative dynamics of LQC are in two folds. First, since there are quantization ambiguities in constructing the
Hamiltonian constraint operator in LQC, it is crucial to check whether the key features of LQC, such as the quantum bounce and effective scenario, are robust against the ambiguities. Second, since LQC serves as a simple
arena to test ideas and constructions induced in the full LQG, it is important to implement those treatments from the full theory to LQC as more as possible. Unlike the usual treatment in spatially flat and homogeneous models, the Euclidean and Lorentz terms have to be quantized separately in full LQG. Therefore, this
kind of quantization procedure which kept the distinction of the Lorentz and Euclidean terms was proposed as alternative dynamics for LQC \cite{YDM2}. It was shown in the resulted canonical effective theory that the classical big bang is again replaced by a quantum bounce. Hence it is desirable to study such kind of prediction from different perspective. Moreover, it is desirable to examine the idea and construction of SFMs by the simplified models of LQC.

The main results of the present paper can be summarized as follows. The path integral formulation is constructed for spatially flat FRW models under the framework of LQC with two alternative dynamics. In both models, we can express the transition amplitude in both the timeless and the deparameterized frameworks by multiple group averaging procedure. We can derive the effective Hamiltonians from both viewpoints. It turns out that in both models the resulted effective dynamics from the timeless and the deparameterized path integrals are equivalent to each other. This indicates the equivalence between two frameworks of path integral. Moreover, the modified Friedmann equations for both models are also obtained and coincide with the corresponding equations in \cite{YDM2} if the higher-order quantum corrections therein are neglected. This indicates the equivalence of the canonical approach and the path integral approach in LQC. Since our path integral approach inherits significant features of SFMs, it provides certain good evidence to support the scheme of SFMs. In both models, the \emph{quantum bounce} will replace the \emph{big bang singularity} due to the modified Friedmann equations when the matter density increase to the magnitude of \emph{Planck density}. Hence the quantum bounce resolution of big bang singularity in LQC is robust against the quantization ambiguities of the Hamiltonian constraint. Moreover, the alternative modified Friedmann equations (\ref{Fdm eq1}) and (\ref{Fdm eq2}) set up new arenas for studying phenomenological issues of LQC.

\section*{ACKNOWLEDGMENTS}

We would like to thank Haiyun Huang for helpful discussion. This work is supported by NSFC (No.10975017) and the Fundamental Research Funds for the Central Universities.

\end{document}